\documentclass[journal=jctc,type=article]{achemso}
\setkeys{acs}{maxauthors=0,articletitle=true}

\usepackage{achemso}
\usepackage{graphics}
\usepackage{amssymb,amsfonts}
\usepackage{graphicx}
\usepackage[table]{xcolor}
\usepackage{multirow}
\usepackage{caption}
\usepackage{booktabs}
\usepackage{amsmath}
\usepackage{amsopn}
\usepackage{siunitx}
\usepackage{bm}
\usepackage{color}
\usepackage{ulem}

\SectionNumbersOn
\DeclareMathOperator{\tr}{Tr}

\author{Janus J. Eriksen}
\email{janus.eriksen@bristol.ac.uk}
\affiliation{School of Chemistry, University of Bristol, Cantock's Close, Bristol BS8 1TS, United Kingdom}
\author{J{\"u}rgen Gauss}
\email{gauss@uni-mainz.de}
\affiliation{Department Chemie, Johannes Gutenberg-Universit{\"a}t Mainz, Duesbergweg 10-14, 55128 Mainz, Germany}

\title[TITLE]{Ground and Excited State First-Order Properties in Many-Body Expanded Full Configuration Interaction Theory}

\begin{document}

%
%%%%%%%%%
%  ABSTRACT  
%%%%%%%%%
%
\begin{abstract}

The recently proposed many-body expanded full configuration interaction (MBE-FCI) method is extended to excited states and static first-order properties different from total, ground state correlation energies. Results are presented for excitation energies and (transition) dipole moments of two prototypical, heteronuclear diatomics---LiH and MgO---in augmented correlation consistent basis sets of up to quadruple-$\zeta$ quality. Given that MBE-FCI properties are evaluated without recourse to a sampled wave function and the storage of corresponding reduced density matrices, the memory overhead associated with the calculation of general first-order properties only scales with the dimension of the desired property. In combination with the demonstrated performance, the present developments are bound to admit a wide range of future applications by means of many-body expanded treatments of electron correlation.

\end{abstract}

\newpage

%
%%%%%%%%%%%
%  INTRODUCTION
%%%%%%%%%%%
%

\section{Introduction}\label{introduction_sect}

The ultimate benchmark of electronic structure theory~\cite{knowles_handy_fci_cpl_1984,knowles_handy_fci_jcp_1989,olsen_fci_jcp_1988,olsen_fci_cpl_1990,olsen_bond_break_h2o_jcp_1996,olsen_bond_break_n2_jcp_2000,sherrill_ch2_jcp_1998,abrams_sherrill_bond_break_c2_jcp_2004,booth_alavi_fciqmc_bond_break_c2_jcp_2011,booth_alavi_fciqmc_nature_2013}, full configuration interaction (FCI) theory has attracted renewed interest over the past decade due to the availability of modern, scalable hardware~\cite{fales_levine_gpu_fci_jcp_2015,vogiatzis_parallel_mcscf_fci_jcp_2017} as well as the emergence of a myriad of new ways in which the exact solution to the electronic Schr{\"o}dinger equation within a given one-electron basis set may be recast~\cite{white_dmrg_prl_1992,white_dmrg_prb_1993,white_martin_dmrg_jcp_1999,mitrushenkov_palmieri_dmrg_jcp_2001,chan_head_gordon_dmrg_jcp_2002,chan_dmrg_jcp_2004,sharma_chan_dmrg_2012,chan_dmrg_review_jcp_2015,guo_chan_pdmrg_jctc_2018,bytautas_ruedenberg_ceeis_jcp_2004,bytautas_ruedenberg_ceeis_jcp_2005,bytautas_ruedenberg_ceeis_jpca_2010,ruedenberg_windus_mbe_jpca_2017,ten_no_fcc_prl_2018,piecuch_monte_carlo_cc_prl_2017,piecuch_monte_carlo_cc_jcp_2018,piecuch_monte_carlo_eom_cc_jcp_2019,piecuch_mg_dimer_sci_adv_2020,zimmerman_ifci_jcp_2017_1,zimmerman_ifci_jcp_2017_2,petruzielo_umrigar_spmc_prl_2012,holmes_umrigar_heat_bath_fock_space_jctc_2016,holmes_umrigar_heat_bath_ci_jctc_2016,sharma_umrigar_heat_bath_ci_jctc_2017,li_sharma_umrigar_heat_bath_ci_jcp_2018,fales_koch_martinez_rrfci_jctc_2018,liu_hoffmann_ici_jctc_2016,liu_hoffmann_ici_jctc_2020,coe_ml_ci_jctc_2018,tubman_whaley_selected_ci_jcp_2016,tubman_whaley_selected_ci_jctc_2020,tubman_whaley_selected_ci_pt_arxiv_2018,loos_cipsi_jcp_2018,schriber_evangelista_selected_ci_jcp_2016,zhang_evangelista_projector_ci_jctc_2016,schriber_evangelista_adaptive_ci_jctc_2017,lu_coord_descent_fci_jctc_2019,berkelbach_fci_fri_jctc_2019,berkelbach_fci_fri_jctc_2020}. The majority of these efforts have focussed on the computation of ground state energies, although studies of the ordering and spacing of excited states have also started to receive attention~\cite{zimmerman_ifci_jpca_2017,holmes_sharma_heat_bath_ci_excited_states_jcp_2017,chien_zimmerman_heat_bath_ci_excited_states_jpca_2017,loos_jacquemin_cipsi_exc_state_jctc_2018,loos_jacquemin_cipsi_exc_state_jctc_2019,loos_jacquemin_cipsi_exc_state_jctc_2020,loos_scemama_jacquemin_cipsi_exc_state_jpcl_2020}. The study of general first-order properties at the near-exact level, on the other hand, has been left somewhat ignored, following its pinnacle in the 1990s~\cite{bauschlicher_taylor_dipole_fci_tca_1987,bauschlicher_langhoff_trans_dipole_fci_tca_1988,cremer_bartlett_dipole_cpl_1993,koch_jorgensen_trans_dipole_ccsd_jcp_1994,halkier_gauss_dipole_fci_jcp_1999,halkier_jorgensen_dipole_cbs_jcp_1999,bak_olsen_dipole_cpl_2000}. Given that excitation energies and molecular multipole moments are physical observables, whereas the ground state energy is not, a strong incentive from the chemical sciences arguably exists in favour of the development of new theoretical tools that may universally allow for the calculation of a wide range of properties in order to complement and guide experimental endeavours. To that end, we note how the simulation of electronic properties at various levels of theory has continued to gather momentum over the years, as evidenced by a number of recent benchmarks on the topic, e.g., assessments of Kohn-Sham density functional theory (KS-DFT) dipole moments~\cite{hickey_rowley_dipole_jpca_2014,verma_truhlar_dipole_pccp_2017,hait_head_gordon_dipole_jctc_2018}. In the present work, however, our focus will differ pronouncedly from such endeavours as we will be strictly concerned with the prediction of first-order properties of near-FCI rather than approximate (truncated) quality.\\

As an example of an exception to the general trend discussed above, quantum Monte Carlo approaches have long retained an interest in the near-exact estimation of general expectation values~\cite{foulkes_mitas_needs_rajagopal_vmc_rev_mod_phys_2001,gaudoin_pitarke_dmc_prl_2007,per_snook_russo_dmc_prb_2012}. Expressed on its usual form as a linear combination of Slater determinants in a discrete basis set, the exact $N$-electron wave function may be described in a systematically improvable fashion by means of initiator FCI quantum Monte Carlo~\cite{booth_alavi_fciqmc_jcp_2009,cleland_booth_alavi_fciqmc_jcp_2010} ($i$-FCIQMC). In the complete absence of initiator bias and in the limit of long samplings, a stochastic sampling of the FCI wave function, in turn to construct the two-electron reduced density matrix (2-RDM), $\Gamma_{pq,rs}$, becomes exact. From the 2-RDM, the one-electron analogue (1-RDM), $\gamma_{pq}$, required for the calculation of, e.g., electronic dipole moments~\cite{thomas_booth_fciqmc_jcp_2015}, may be trivially obtained through a partial trace operation. In practice, the so-called replica trick~\cite{overy_alavi_fciqmc_jcp_2014} is employed whereby two independent, randomly initialized FCIQMC calculations are performed for a given state to arrive at significantly less biased 2-RDMs~\cite{booth_alavi_tew_fciqmc_jcp_2012}, using coefficients from both replica runs. The calculation of transition RDMs, from which transition dipole moments are computed, is more involved, but may these days also be realized by FCIQMC~\cite{booth_alavi_fciqmc_jcp_2017}.\\

Recently, many-body expanded FCI~\cite{eriksen_mbe_fci_jpcl_2017,eriksen_mbe_fci_weak_corr_jctc_2018,eriksen_mbe_fci_strong_corr_jctc_2019,eriksen_mbe_fci_general_jpcl_2019} (MBE-FCI) theory has been proposed by us as an alternative to selected CI~\cite{malrieu_cipsi_jcp_1973,harrison_selected_ci_jcp_1991,stampfuss_wenzel_selected_ci_jcp_2005}, operating instead by performing an orbital-based decomposition of the FCI correlation problem. By solving for correlation energies without recourse to the full $N$-electron electronic wave function, the MBE-FCI method has emerged as an accurate approximation to exact theory, applicable in extended basis sets~\cite{eriksen_mbe_fci_strong_corr_jctc_2019} and for moderate-sized molecular system~\cite{eriksen_benzene_jpcl_2020}. Specifically, a strict partitioning is enforced of the complete set of molecular orbitals (MOs) of a system into a reference and an expansion space. A complete active space CI (CASCI) calculation is then performed in the former of these two spaces, while the residual correlation in the latter space is recovered by an orbital-based MBE. In the present work, we will report on an extension of MBE-FCI theory to excited states and first-order properties for arbitrary states with an aim at computing near-exact properties in a unified framework on par with alternatives from the literature, e.g., $i$-FCIQMC theory.

%
%%%%%%%
%  THEORY
%%%%%%%
%

\section{Theory}\label{theory_sect}

In the MBE-FCI method, the FCI correlation energy is decomposed as
\begin{align}
E^{0}_{\text{FCI}} = E^{0}_{\text{ref}} + \sum_{p}\epsilon^{0}_{p} + \sum_{p<q}\Delta\epsilon^{0}_{pq} + \sum_{p<q<r}\Delta\epsilon^{0}_{pqr} + \ldots \label{mbe_corr_energy_eq}
\end{align}
In Eq. \ref{mbe_corr_energy_eq}, the MOs of the expansion space (of size $M_{\text{exp}}$) of unspecified occupancy are labelled by generic indices $\{p,q,r,s,\ldots\}$, and $\epsilon^{0}_{p}$ designates the correlation energy of a CASCI calculation for the ground state (0) in the composite space of orbital $p$ and all of the reference space MOs. For a general tuple of $m$ MOs, $[\Omega]_{m}$, its $m$th-order increment, $\Delta\epsilon_{[\Omega]_{m}}$, is defined through recursion. As outlined in Ref. \citenum{eriksen_benzene_jpcl_2020}, MOs are screened away from the full expansion space at each order according to their relative (absolute) magnitude, which in turn results in a reduced number of increment calculations at the orders to follow. Ultimately, these successive screenings lead to the convergence of an MBE-FCI calculation. In analogy with Eq. \ref{mbe_corr_energy_eq}, excitation energies may now be computed by an expansion of the energetic gap between the ground and an excited state, $E^{0n}$, rather than the correlation energy
\begin{align}
E^{0n}_{\text{FCI}} = E^{0n}_{\text{ref}} + \sum_{p}\epsilon^{0n}_{p} + \sum_{p<q}\Delta\epsilon^{0n}_{pq} + \sum_{p<q<r}\Delta\epsilon^{0n}_{pqr} + \ldots \label{mbe_ex_energy_eq}
\end{align}
As a CASCI calculation in an active space absent of electron correlation yield no correlation energy (comprising only the Hartree-Fock (HF) solution) and hence no excited states, $\epsilon^{0n}_{p}$ in Eq. \ref{mbe_ex_energy_eq} will be defined on par with $\epsilon^{0}_{p}$ in Eq. \ref{mbe_corr_energy_eq}. For the calculation of static properties, we will here exemplify how this may be achieved in the context of MBE-FCI theory by focussing on electronic dipole and transition dipole moments. From the wave function coefficients of an individual CASCI calculation, the corresponding 1-RDM, $\bm{\gamma}^{n}$, for state $n$ may be readily computed, from which an electronic dipole moment is given as
\begin{align}
\bm{\mu}^{n}_{p} = -\sum_{r}\tr[\bm{\mu}_{r}\bm{\gamma}^{n}] \label{dipmom_eq}
\end{align}
in terms of dipole integrals, $\bm{\mu}_{r}$, in the MO basis for each of the three cartesian components. Letting these quantities take up the role of correlation or excitation energies in Eqs. \ref{mbe_corr_energy_eq} and \ref{mbe_ex_energy_eq}, respectively, results in the following decomposition of the FCI electronic dipole moment
\begin{align}
\bm{\mu}^{n}_{\text{FCI}} = \bm{\mu}^{n}_{\text{ref}} + \sum_{p}\bm{\mu}^{n}_{p} + \sum_{p<q}\Delta\bm{\mu}^{n}_{pq} + \sum_{p<q<r}\Delta\bm{\mu}^{n}_{pqr} + \ldots \label{mbe_dipole_eq}
\end{align}
Adding the nuclear component, $\bm{\mu}_{\text{nuc}} = \sum_{K}Z_{K}\bm{r}_{K}$, returns the molecular dipole moment. Being a vector rather than a scalar quantity, the screening procedure proceeds along all three cartesian components ($x,y,z$) in the case of dipole moments and must be simultaneously fulfilled for all if a given MO is to be screened away from the expansion space.\\

For the calculation of ground state dipole moments, increments are evaluated against the HF dipole moment---in line with the correlation energies entering Eq. \ref{mbe_corr_energy_eq}---while excited state electronic dipole moments are evaluated in the absence of a zero point (as in Eq. \ref{mbe_ex_energy_eq}). Finally, transition dipole moments, $\bm{t}^{0n}$, are evaluated on par with Eq. \ref{mbe_dipole_eq}, except for the fact that the individual increments are computed on the basis of transition 1-RDMs, $\bm{\bm{\gamma}}^{0n}$, which may be arrived at using the wave functions of both states involved in a given CASCI calculation.

%
%%%%%%%%%%%
%  COMP. DETAILS
%%%%%%%%%%%
%

\section{Computational Details}\label{comp_details_sect}

In comparison with MBE-FCI for the computation of ground state correlation energies, the overhead in terms of added compute time and memory demands is minimal, also when contrasted with $i$-FCIQMC and related methods that sample the full 2-RDM, retrieve the 1-RDM, before contracting this with the appropriate property integrals. By virtue of the small active spaces involved in all individual CASCI increment calculations of an MBE-FCI expansion, the additional cost associated with the computation of $\bm{\gamma}^{n}$ or $\bm{\gamma}^{0n}$, depending on the property in question, is negligible, as we will show below. With regards to storage requirements, these are the same for correlation and excitation energies, while for (transition) dipole moments the only difference is that the increment quantities to store are now tensorial, rather than scalar---specifically for $\bm{\mu}^{n}$ and $\bm{t}^{0n}$, the memory requirements increase by a factor 3 over an MBE-FCI calculation of the correlation energy. All results to follow have been obtained in an embarrassingly parallel manner using the open-source {\texttt{PyMBE}} code~\cite{pymbe}, which in turn employs the {\texttt{PySCF}} code~\cite{pyscf_prog,pyscf_wires_2018,pyscf_jcp_2020} for all electronic structure kernels. All calculations of the present work, for which timings are reported, were run on Intel Xeon E5-2697v4 (Broadwell) nodes (36 cores $@$ 2.3 GHz, 128 GB).

%
%%%%%%%%
%  RESULTS
%%%%%%%%
%

\section{Results}\label{results_sect}

\begin{figure}[ht!]
\begin{center}
\includegraphics[width=.9\textwidth]{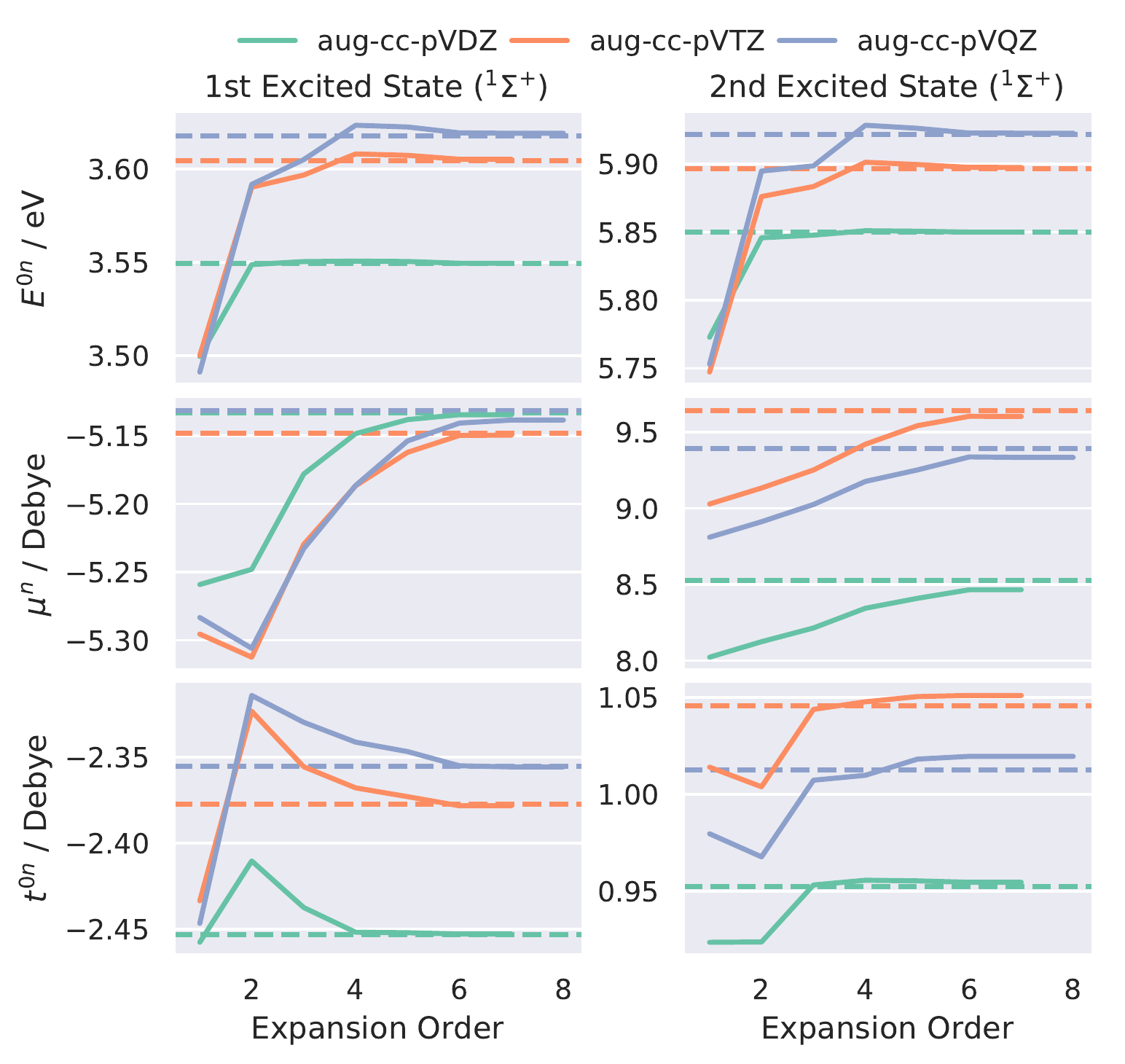}
\caption{Excitation energies ($E^{0n}$, upper panels), dipole moments ($\mu^{n}$, center panels), and transition dipole moments ($t^{0n}$, lower panels) for the first two excited states (${^{1}}\Sigma^{+}$ symmetry) of LiH in the aug-cc-pV$X$Z basis sets. Solid and dashed lines denote MBE-FCI and $i$-FCIQMC~\cite{booth_alavi_fciqmc_jcp_2017} results, respectively.}
\label{results_lih_fig}
\end{center}
\vspace{-0.6cm}
\end{figure}
Figure \ref{results_lih_fig} presents results for the first and second excited states (${^{1}}\Sigma^{+}$ symmetry) of LiH in the aug-cc-pV$X$Z basis sets~\cite{dunning_1_orig,dunning_4_aug}, with MBE-FCI results for excitation energies, dipole moments, and transition dipole moments compared to corresponding $i$-FCIQMC results from Ref. \citenum{booth_alavi_fciqmc_jcp_2017}. Calculations for a given state and a given property were run independently. The MBE-FCI calculations have all been run in $C_{2\text{v}}$ symmetry using the MOs of a state-averaged complete active space self-consistent field~\cite{roos_casscf_acp_1987} (CASSCF) calculation in an ($4e$,$7$o) active space coinciding with the employed reference space, and the screening thresholds are discussed in the Supporting Information (SI). The results in Figure \ref{results_lih_fig} and Figure S1 of the SI collectively show how the performance of MBE-FCI for ground state correlation energies is reflected not only in a corresponding accuracy for ground state dipole moments, but also transfers to excited state properties as well. In general, the convergence profiles in the 6 individual plots of Figure \ref{results_lih_fig} are all different, as are the MO manifolds being screened away in the calculations, attesting to the facts that the properties in question are inherently unrelated and that the MBE-FCI method is flexible enough to cope with this in an orbital expansion framework.\\

\vspace{-.4cm}
\begin{table}[ht]
\begin{center}
\begin{tabular}{l|r|r|r}
\toprule
\multicolumn{1}{c|}{\multirow{2}{*}{Basis Set}} & \multicolumn{3}{c}{Property} \\
& \multicolumn{1}{c|}{$E^{0n}$} & \multicolumn{1}{c|}{$\mu^{n}$} & \multicolumn{1}{c}{$t^{0n}$} \\
\midrule\midrule
\multicolumn{4}{c}{Ground State (${^{1}}\Sigma^{+}$)} \\
\midrule
aug-cc-pVDZ & $3.9$ & $4.9$ & \multicolumn{1}{c}{---} \\
aug-cc-pVDZ ($\pi$) & $0.1$ & $0.1$ & \multicolumn{1}{c}{---} \\
aug-cc-pVTZ ($\pi$) & $10.2$ & $11.6$ & \multicolumn{1}{c}{---} \\
aug-cc-pVQZ ($\pi$) & $443.4$ & $465.1$ & \multicolumn{1}{c}{---} \\
\midrule
\multicolumn{4}{c}{1st Excited State (${^{1}}\Sigma^{+}$)} \\
\midrule
aug-cc-pVDZ & $10.6$ & $12.9$ & $12.9$ \\
aug-cc-pVDZ ($\pi$) & $0.3$ & $0.3$ & $0.3$ \\
aug-cc-pVTZ ($\pi$) & $16.6$ & $19.4$ & $19.4$ \\
aug-cc-pVQZ ($\pi$) & $866.2$ & $985.6$ & $980.5$ \\
\midrule
\multicolumn{4}{c}{2nd Excited State (${^{1}}\Sigma^{+}$)} \\
\midrule
aug-cc-pVDZ & $10.1$ & $12.8$ & $12.9$ \\
aug-cc-pVDZ ($\pi$) & $0.2$ & $0.3$ & $0.3$ \\
aug-cc-pVTZ ($\pi$) & $14.5$ & $16.4$ & $17.2$ \\
aug-cc-pVQZ ($\pi$) & $813.4$ & $904.9$ & $897.3$ \\
\bottomrule
\end{tabular}
\vspace{0.2cm}
\caption{Total timings (core hours) for the LiH calculations in Figure \ref{results_lih_fig}.${^{a,b}}$}
\label{timings_lih_table}
{\footnotesize{${^{a}}$ Intel Xeon E5-2697v4 (Broadwell) nodes (36 cores $@$ 2.3 GHz, 128 GB).}}\\
{\footnotesize{${^{b}}$ ($\pi$) indicates $\pi$-pruning.}}
\vspace{-0.5cm}
\end{center}
\end{table}
All of the results of the present work have been obtained using the $\pi$-pruning of Ref. \citenum{eriksen_mbe_fci_strong_corr_jctc_2019} (generalized to $C_{\infty\text{v}}/C_{2\text{v}}$ point groups), which is a prescreening filter that prunes away all increment calculations that fail to simultaneously include the $x$- and $y$-components of a given pair of degenerate $\pi$-orbitals. The use of this $\pi$-pruning filter results in much shorter (faster) expansions for linear molecules belonging to non-Abelian point groups, while at the same time warranting convergence onto states spanned by the correct irreducible representation (${^{1}}\Sigma^{+}/A_1$ in our case). Table \ref{timings_lih_table} presents timings in units of core hours that clearly show not only the minimal overhead associated with computing excited and non-energetic properties, but also the reduction in compute time that results from the use of $\pi$-pruning (indicated by ($\pi$) in the table). However, in order to obtain accurate results across all of the tested properties (e.g., for the second root), a tighter screening procedure is needed in combination with $\pi$-pruning, cf. Figs. S2 and S3 of the SI. Please note that the timings in Table \ref{timings_lih_table} correspond to executions on a single core; in practice, MBE-FCI calculations are performed on all cores across a number of nodes. For instance, using two of the Broadwell nodes described in Section \ref{comp_details_sect}, the timings in Table \ref{timings_lih_table} are reduced by roughly a factor of 72.\\

\begin{figure}[ht!]
\begin{center}
\includegraphics[width=.9\textwidth]{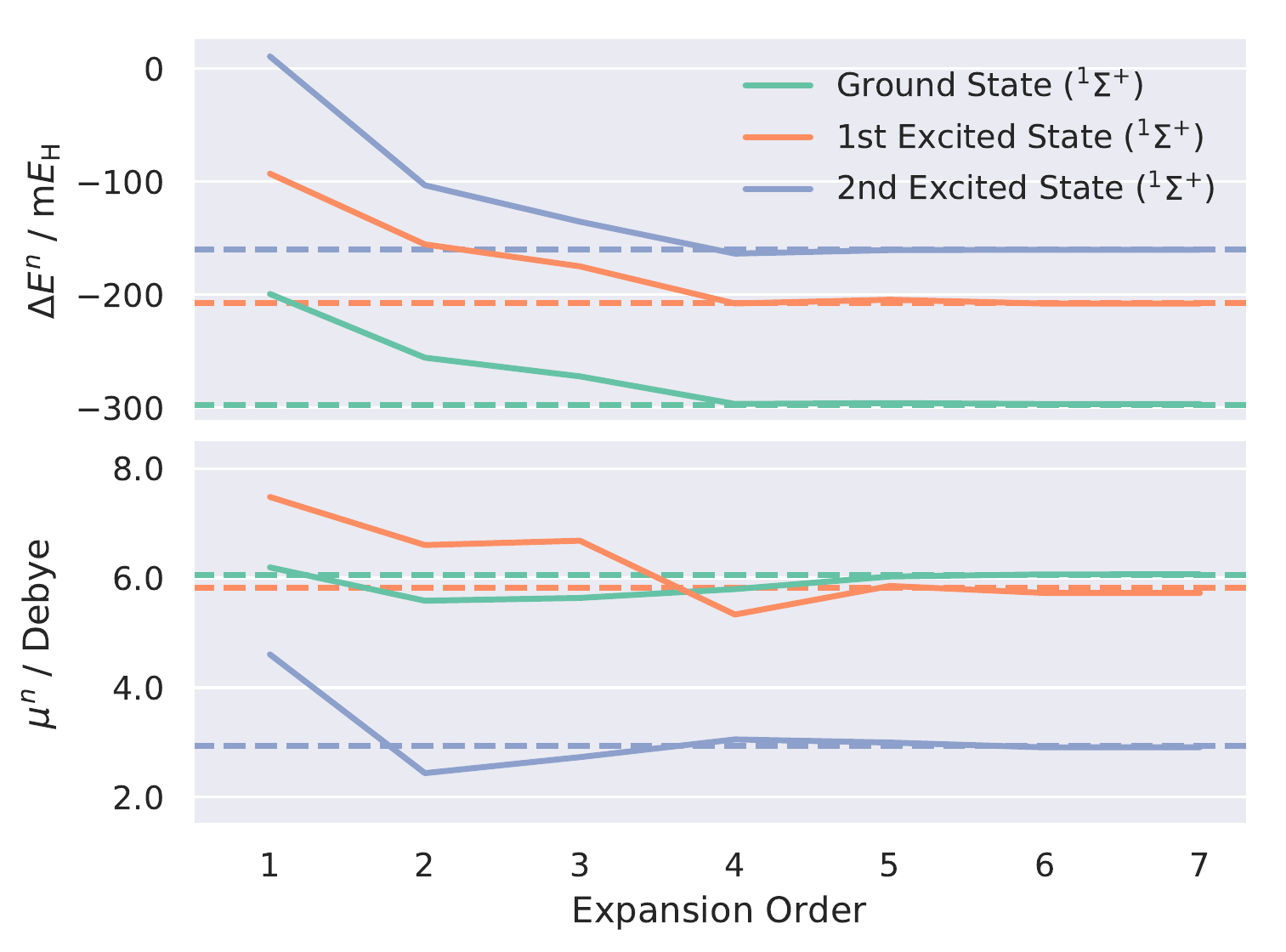}
\caption{Correlation energies ($\Delta E^{n}$, upper panel) and dipole moments ($\mu^{n}$, lower panel) for the ground and first two excited states (${^{1}}\Sigma^{+}$ symmetry) of MgO in the aug-cc-pVDZ basis set. Solid and dashed lines denote MBE-FCI and $i$-FCIQMC~\cite{booth_alavi_fciqmc_jcp_2017} results, respectively.}
\label{results_mgo_fig}
\end{center}
\vspace{-0.6cm}
\end{figure}
We next turn to the problem of MgO in an aug-cc-pVDZ basis set, for which the frozen-core FCI correlation problem is described by the distribution of 16 electrons in 48 orbitals. MBE-FCI results---obtained using reference spaces spanned by (state-averaged) CASSCF($8e$,$8$o) calculations---are presented in Figure \ref{results_mgo_fig}, again in comparison with $i$-FCIQMC results from Ref. \citenum{booth_alavi_fciqmc_jcp_2017}. In contrast to the earlier LiH example, the $i$-FCIQMC results for MgO are noted in Ref. \citenum{booth_alavi_fciqmc_jcp_2017} to be somewhat less converged and hence associated with an increased degree of uncertainty (particularly for $\bm{\mu}^{1}$). Be that as it may, $i$-FCIQMC was still deemed more accurate than, e.g., high-level coupled cluster results~\cite{ccsdt_paper_1_jcp_1987,*ccsdt_paper_1_jcp_1988_erratum,ccsdt_paper_2_cpl_1988} (CCSDT), in particular in the case of excited state dipole moments, which from the stochastic wave functions in Ref. \citenum{booth_alavi_fciqmc_jcp_2017} were shown to depend crucially on highly excited determinants which are less successfully described by means of CCSDT. For instance, CCSDT differs from $i$-FCIQMC by $+0.04$, $-0.71$, and $-0.78$ Debye in the prediction of the three dipole moments in question~\cite{booth_alavi_fciqmc_jcp_2017}.\\

From the MBE-FCI results in Figure \ref{results_mgo_fig} (all obtained using $\pi$-pruning and the same screening thresholds as for LiH), we observe how the method correctly converges onto the individual states of interest. Despite the fact that the three states lie in close proximity of each other, MBE-FCI succeeds in distinguishing between them. In relation to the troublesome dipole moment for the first excited state, this is also the result in Figure \ref{results_mgo_fig} for which the largest discrepancy with respect to $i$-FCIQMC is observed. In order to determine a more accurate value of $\bm{\mu}^{1}$, more walkers would need be employed in $i$-FCIQMC and a more conservative screening protocol would need be used in the context of MBE-FCI (as for LiH in Figure S3 of the SI). Finally, consumed core hours are presented in Table \ref{timings_mgo_table}. As for LiH, these results once again convincingly illustrate the efficacy of the MBE-FCI method as well as the low penalty associated with computing dipole moments over correlation and excitation energies.\\
\vspace{-0.5cm}
\begin{table}[ht]
\begin{center}
\begin{tabular}{l|r|r}
\toprule
\multicolumn{1}{c|}{\multirow{2}{*}{State}} & \multicolumn{2}{c}{Property} \\
& \multicolumn{1}{c|}{$E^{n}$} & \multicolumn{1}{c}{$\mu^{n}$} \\
\midrule\midrule
0 (${^{1}}\Sigma^{+}$) & 4,412 & 6,566 \\
1 (${^{1}}\Sigma^{+}$) & 11,320 & 16,720 \\
2 (${^{1}}\Sigma^{+}$) & 24,026 & 36,497 \\
\bottomrule
\end{tabular}
\vspace{0.2cm}
\caption{Total timings (core hours) for the MgO calculations in Figure \ref{results_mgo_fig}.${^{a}}$}
\label{timings_mgo_table}
{\footnotesize{${^{a}}$ Intel Xeon E5-2697v4 (Broadwell) nodes (36 cores $@$ 2.3 GHz, 128 GB).}}
\vspace{-0.5cm}
\end{center}
\end{table}
%

%
%%%%%%%%
%  SUMMARY
%%%%%%%%
%

\vspace{-.4cm}
\section{Summary and Conclusions}\label{summary_sect}

In summary, we have reported on a new extension of MBE-FCI theory to molecular first-order properties, valid for both ground and excited states. The performance of the resulting implementation has been verified through calculations on the first three roots of the LiH and MgO diatomics by comparing our results to state-of-the-art $i$-FCIQMC. On the basis of the proven accuracy and efficacy, we foresee that MBE-FCI has the potential to act as a new near-exact benchmark method for first-order properties of small- to modest-sized molecular systems in the years to come. Although the theory depends on a choice of reference space that necessarily encompasses the main determinant(s) of the target state, we are currently working on automatic selection procedures that will allow for the black-box division of a system's total set of MOs into optimal reference and expansion spaces.

%
%%%%%%%%%%%%%%%
%  ACKNOWLEDGMENTS 
%%%%%%%%%%%%%%%
%
\section*{Acknowledgments}

J.J.E. is grateful to the Alexander von Humboldt Foundation as well as the Independent Research Fund Denmark for financial support. The authors furthermore acknowledge access awarded to the Galileo supercomputer at CINECA (Italy) through the $18^{\text{th}}$ PRACE Project Access Call and the Johannes Gutenberg-Universit{\"a}t Mainz for computing time granted on the MogonII supercomputer.

%
%%%%%%%%%%%%%%%%%%%%%%%%%%%%%%%%%%%%%%%%%%%%%%%%%%%%%%%%%%%%%%%%%%%
%                                                                    			     Supporting Information
%%%%%%%%%%%%%%%%%%%%%%%%%%%%%%%%%%%%%%%%%%%%%%%%%%%%%%%%%%%%%%%%%%%
%
\section*{Supporting Information}

All results have been tabulated in the Supporting Information, cf. Tables S1 and S2. Additional results are presented for the ground state of LiH (Figure S1), for its excited states in the aug-cc-pVDZ basis set in the absence of $\pi$-pruning (Figure S2), and for its second excited state using tighter screening thresholds (Figure S3).

%
%%%%%%%%%%%%%%%%%%%%%%%%%%%%%%%%%%%%%%%%%%%%%%%%%%%%%%%%%%%%%%%%%%%
%                                                                    			     Data Availability
%%%%%%%%%%%%%%%%%%%%%%%%%%%%%%%%%%%%%%%%%%%%%%%%%%%%%%%%%%%%%%%%%%%
%
\section*{Data Availability}

Data in support of the findings of this study are available within the article and its SI.

\newpage

\providecommand{\latin}[1]{#1}
\makeatletter
\providecommand{\doi}
  {\begingroup\let\do\@makeother\dospecials
  \catcode`\{=1 \catcode`\}=2 \doi@aux}
\providecommand{\doi@aux}[1]{\endgroup\texttt{#1}}
\makeatother
\providecommand*\mcitethebibliography{\thebibliography}
\csname @ifundefined\endcsname{endmcitethebibliography}
  {\let\endmcitethebibliography\endthebibliography}{}

\end{document}